# Diversity in Ranking using Negative Reinforcement

Rama Badrinath
Department of Computer Science and Automation
Indian Institute of Science
Bangalore, India
ramab@csa.iisc.ernet.in

C E Veni Madhavan
Department of Computer Science and Automation
Indian Institute of Science
Bangalore, India
cevm@csa.iisc.ernet.in

## ABSTRACT
In this paper, we consider the problem of diversity in ranking of the nodes in a graph. The task is to pick the top-k nodes in the graph which are both 'central' and 'diverse'. Many graph-based models of NLP like text summarization, opinion summarization involve the concept of diversity in generating the summaries. We develop a novel method which works in an iterative fashion based on random walks to achieve diversity. Specifically, we use negative reinforcement as a main tool to introduce diversity in the Personalized PageRank framework. Experiments on two benchmark datasets show that our algorithm is competitive to the existing methods.

## 1. INTRODUCTION
Ranking of data items represented in the form of a graph has many applications. Consider the example of a co-author network. The goal is to pick the most influential authors from different areas. A simple approach would be to use a centrality measure like PageRank [2], and then to pick the top scoring vertices in the graph. However, top ranking authors might not cover all the different areas, because top ranking nodes would be very near to each other (redundancy). Hence, we have to consider both 'centrality' and 'diversity' in the ranking process. A straight forward approach to diversity is to first cluster the graph and the pick the cluster centres. The main drawback is that we may not know the number of clusters prior to clustering.

The concept of diversity is used in a variety of applications like Text summarization [3], Recommendation systems [14], Query suggestions [9], Search results diversification [1, 5, 12], etc. Before presenting the current approaches to diversity, we will briefly describe the two classic algorithms used to rank nodes in a graph.

The PageRank algorithm [2] tries to find out the central nodes in the graph by using the concept of random walk with teleporting. The score of a node depends on the score of its neighbors. From a vertex $s$, the walk either 'teleports' to a random vertex in the graph with probability $(1-\lambda)$, or with probability $\lambda$ jumps to one of the neighbours of $s$.

$$p(s) = \frac{(1-\lambda)}{N} + \lambda \sum_{v \in V} \frac{sim(s,v)}{\sum_{z \in V} sim(z,v)} p(v) \tag{1}$$

The Personalized PageRank algorithm [6] goes one step ahead and uses the prior distribution for the nodes in the ranking process. Hence, nodes with a high prior value will be favored in case of random jump.

$$p(s) = (1-\lambda)\frac{p^*(s)}{\sum_{z \in V} p^*(z)} + \lambda \sum_{v \in V} \frac{sim(s,v)}{\sum_{z \in V} sim(z,v)} p(v) \tag{2}$$

where, $p^*$ is the prior distribution.

Top-k nodes returned by both the algorithms will be central but not necessarily diverse.

## 2. RELATED WORK
The current methods for diversity fall into two categories:

- **Greedy selection methods**: Iteratively selects one vertex at a time such that the next vertex is central, and as diverse as possible from the set of ranked (picked) vertices.
- **Unified ranking methods**: Both 'centrality' and 'diversity' are considered simultaneously in the ranking process, which runs only once.

Maximum Marginal Relevance (MMR) [3] works in a greedy fashion. The main idea is that the next vertex to be picked must be most prestigious. At the same time, it must be least similar to the set of ranked nodes.

$$MMR \stackrel{\text{def}}{=} \max_{v_i \in R \setminus S} [\lambda Sim(v_i, Q) - (1-\lambda) \max_{v_j \in S} Sim(v_i, v_j)] \quad (3)$$

where, $R$ denotes the set of all nodes, and $S$ is the set of already selected nodes. $Q$ denotes the query given to us and $sim(v_i, Q)$ is the similarity of node $v_i$ to query. Since, we dot not have the notion of query in the current problem, we consider $sim(v_i, Q)$ to be the PageRank score of $v_i$. Hence, the first ranked node will be the node having highest PageRank score. The next set of nodes are picked according to Equation 3.

GRASSHOPPER [13] is another greedy selection method which uses the concept of 'absorbing random walk'. The first ranked node will be the node having highest PageRank score. Next, the ranked nodes are considered as absorbing states, and the time to absorption for each node is calculated. The node with the maximum number of visits is selected as the next ranked item.

DivRank [10] is an unified ranking algorithm. It leverages 'vertex reinforced random walk' to deploy the rich-gets-richer mechanism. The idea is to use time dependent transition probabilities such that the high ranked nodes absorb scores from neighboring nodes. Finally, cluster centres win in the process and get high score.

$$p_T(u, v) = (1 - \lambda) \cdot \frac{p^*(v)}{\sum_{z \in V} p^*(z)} + \lambda \frac{p_0(u, v) \cdot N_T(v)}{D_T(u)} \quad (4)$$

$$D_T(u) = \sum_{v \in V} p_0(u, v) N_T(v) \quad (5)$$

where, $p_0(u, v)$ is the transition probability prior to any reinforcement. $N_T(v)$ is the number of times the walk has visited the node '$v$' up to time $T$.

In the next section, we propose an algorithm named Negative Reinforcement Ranking (NR$^2$), which is basically a greedy selection method. Section 3 discusses the experiments conducted on the benchmark datasets. After a discussion on running time analysis in Section 4, we conclude the paper in Section 5 with some future work.

# 3. NEGATIVE REINFORCEMENT RANKING (NR$^2$)

We borrow the concept of negative reinforcement in random walks from PNR$^2$ [7]. The full details of our algorithm are as follows. Our algorithm is similar in structure to GRASSHOPPER, and picks a diverse set of nodes one at a time. To find the first ranked item, we run the standard Personalized PageRank [6] (PPR) on the graph and pick the top scoring node. Let $W$ denote a row normalized similarity matrix i.e., $\sum_{j=1}^{n} W(i, j) = 1$. Let $r$ denote a preference vector (prior distribution) of the vertices. Then the PPR vector $p$ is given by:

$$p = (1 - \lambda)r + \lambda W' p \quad (6)$$

where, $\lambda$ is the damping factor. Usually, the *power method* is used to solve the above equation to find the PPR vector. However, it can be written as follows:

$$(I - \lambda W')p = (1 - \lambda)r \quad (7)$$

We solve this system of linear equations to find the score vector $p$. As described earlier, we pick the top scoring vertex as the first ranked item. Now, we are ready to select the next set of nodes.

It is crucial that the next item must be very far from the first ranked item. Further it must be central in the remaining subgraph. Hence, we flip the preference value (prior ranking) of the ranked item from positive to negative value and run the PPR again. The effect of this is that the ranked item's negative score tries to pull down the score of its nearby vertices and so on. However, the nodes which are very far from the ranked nodes will not affected. Hence they still get positive reinforcement from their neighbors. The top scoring node denotes the 'central node' in the remaining graph. Therefore, we pick the top scoring node as the next ranked item. Due to negative reinforcement, nodes which are near the ranked item are removed from the race.

After picking the second item, we set the prior ranking of both the ranked items to negative value, and run PPR. The process is continued till top-k nodes are picked. In order to have a good control over the distribution of negative mass from the set of ranked nodes to the set of unranked nodes, we augment the original graph with a new node (having a self edge) called 'absorbing node'. It absorbs positive mass from the set of unranked nodes. This leads in a negative mass propagation to a larger portion of the graph. However, an absorbing node is never picked into the set of ranked nodes.

**Input:** $W$, $r$, $\lambda$, $k$
**Output:** List of top-k nodes '**g**'

1. $A = \emptyset$, $B = \{v_1..v_n\}$
2. Augment the graph with a new node '**d**' (absorbing node) with a self edge.
3. Solve Eqn 7 with $r$ as the prior distrubution. Pick the first ranked item, $g_1$ = argmax$_i$ $p_i$.
   $A = A \cup g_1$, $B = B \setminus g_1$,
4. Repeat: $j = 2$ till $k$
   (a) $r^* = r$
   Normalize $r^*(A)$, $r^*(B)$ seperately.
   (b) $r^*(d) = \beta$
   $r^*(A) = (-\alpha)\, r^*(A)$
   $r^*(B) = (1 + \alpha - \beta)\, r^*(B)$
   (c) Solve Equation 7 with $r^*$ as the prior distribution.
   (d) Pick the next ranked item
   $g_j$ = argmax$_i$ $p_i$.
   $A = A \cup g_j$, $B = B \setminus g_j$,

**The Negative Reinforcement Ranking (NR$^2$) algorithm**

Let '$A$' denote the set of ranked nodes, and '$B$' be the set of unranked nodes. Similarly, let $r(A)$ denote the indices of ranked nodes, and $r(B)$ denote the indices of unranked nodes. The complete algorithm is described above. $\alpha$ and $\beta$ are positive parameters to be chosen. Step 4.b makes sure that $\sum_i r_i = 1$, and hence solution to Equation 7 always exists.

We note that only the preference vector $r^*$ changes in every iteration. The term $(I - \lambda W')$ remains same throughout the process. Therefore we use the LU decomposition method to solve Equation 7. The advantage of this method is that the LU decomposition of the matrix $(I - \lambda W')$ (maximum time taking operation) is done only once. Hence, Step 4.c of the algorithm runs very fast[1].

# 4. EXPERIMENTS

## 4.1 Social Network Analysis

We first test our method on the 'actor social network' introduced in [13]. The dataset consists of 3452 actors from 1027 comedy movies collected from Internet Movie Database[2]. An undirected graph is created by considering actors as nodes. Edge weights are proportional to the number of movies the two actors have acted together in. Self edges are added with unit weight.

The task is to extract top-k actors who are prominent in the graph, taking diversity into account. We expect top actors to be comedians from different countries. This forces diversity in the top actors. This is based on the fact that co-star connections within same country are more likely, thus creating communities within countries.

The diversity quality is measured in terms of the number of unique countries covered - *country coverage*, and the number of unique movies covered - *movie coverage* by the top-k actors. A good diversity based algorithm should maximize both country coverage and movie coverage, since the top-k actors come from different regions. We also use the *density* measure introduced in [10]. Density of a graph is defined as the number of edges present in the network divided by the maximal possible number of edges in the network. Formally it is defined as,

$$d(G) = \frac{\sum_{u \in V} \sum_{v \in V, u \neq v} \mathbf{I}[w(u,v) > 0]}{|V| \times (|V| - 1)} \tag{8}$$

where, $\mathbf{I}(X)$ is an indicator function which returns 1, if the statement $X$ is true and zero otherwise. Intuitively, density of the subgraph created by the top-k actors must be very low, because the top-k vertices must be as independent as possible from one another.

We compare our algorithm with the algorithms PageRank, Personalized PageRank, MMR, Grasshopper and DivRank. In order to find out the optimal parameters for all the algorithms, we divided the actor social graph into two subgraphs. Nodes are picked randomly into two sets, intra-set edges are retained and inter-set edges are discarded. The first subgraph created had 1703 nodes and was used for training and the second subgraph having 1749 nodes was used for testing.

[1]A Matlab implementation of the algorithm is available at `https://sites.google.com/site/iiscrama/home/nr2-ranking`

[2]http://www.imdb.com/

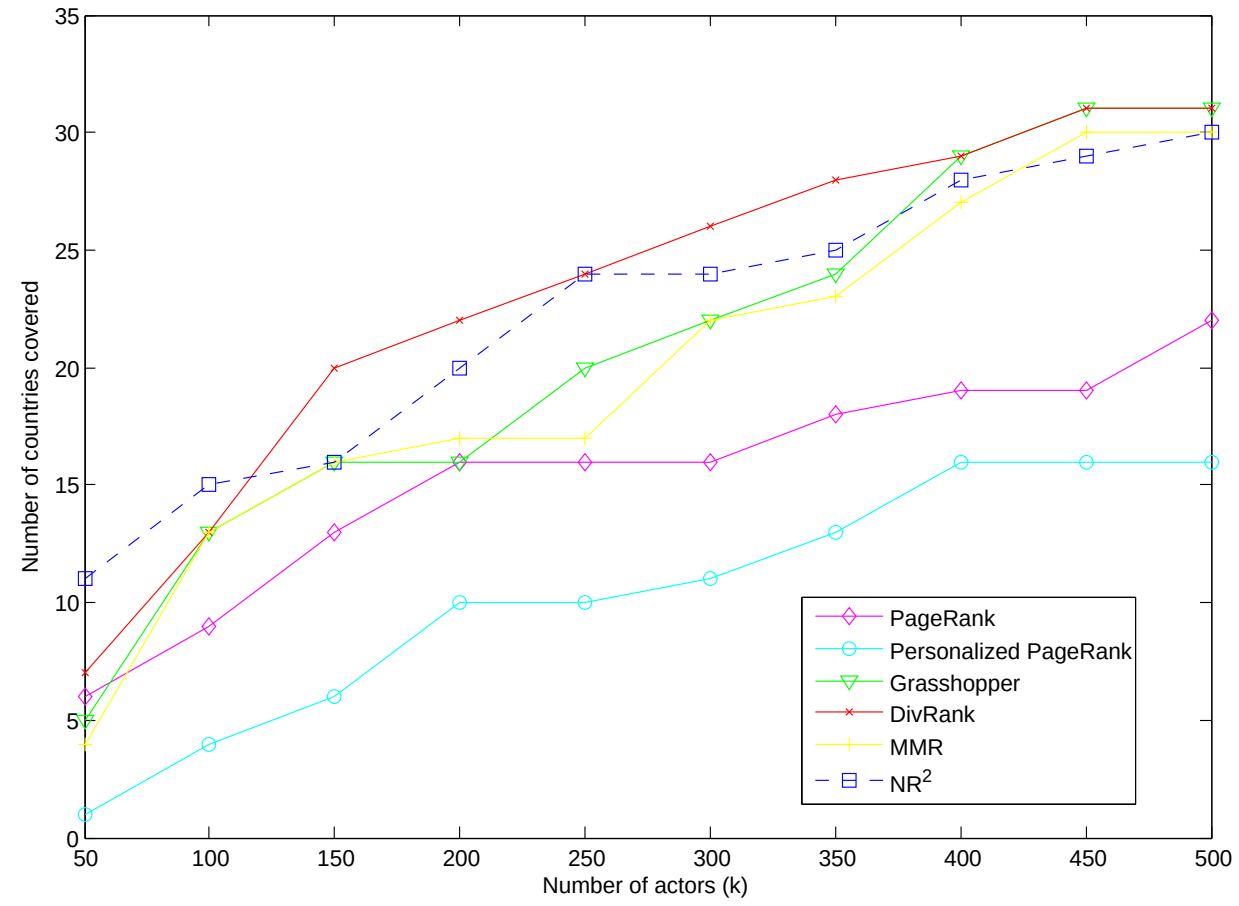

Figure 1: Country coverage.

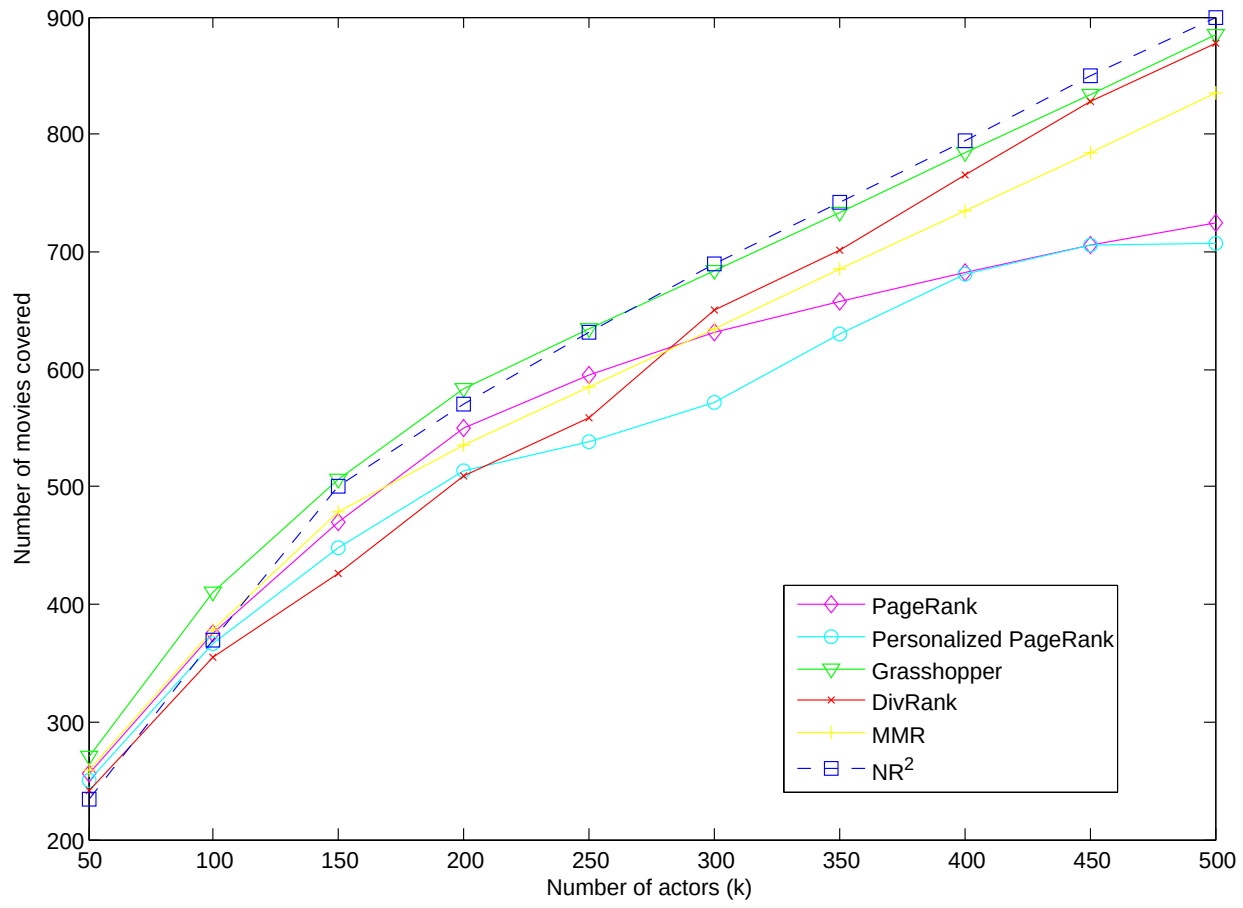

Figure 2: Movie coverage.

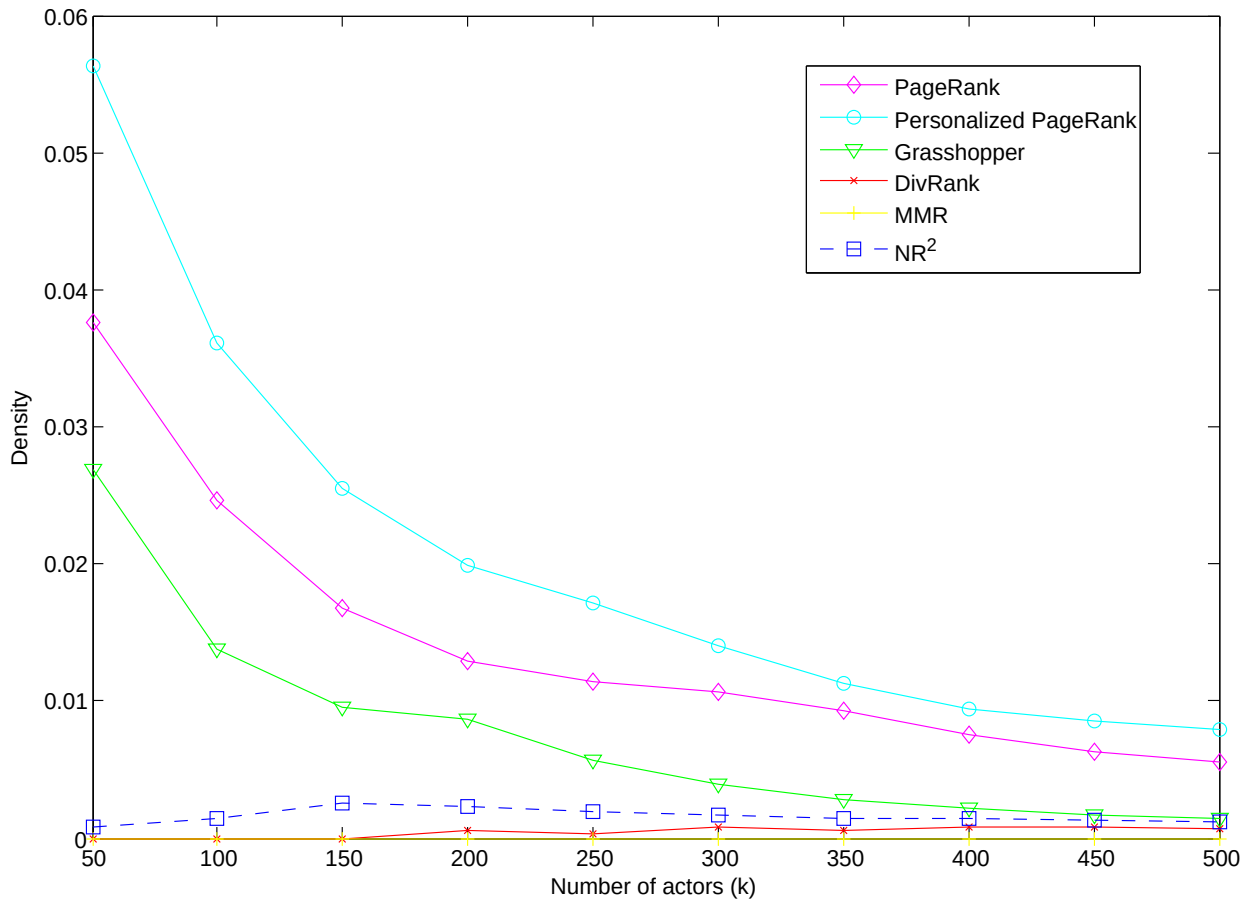

Figure 3: Density (the lower the better).

Fig 1, 2, 3 shows the performance of different algorithms. We see that NR$^2$ performs very well in country coverage, especially when $k$ is small ($k$<100). This is a good property, since in many applications only a few top results are considered. It is also important to note that the density of our algorithm is very less compared to Grasshopper.

The list of top 10 actors returned by NR$^2$ algorithm on the test set is shown in the table below.

| |
|---|
| Eddie Murphy |
| Anthony Anderson |
| Johny Knoxville |
| Luke Wilson |
| Steve Martin |
| Steve Buscemi |
| Breckin Meyer |
| Jackie Chan |
| Brittany Murphy |
| Til Schweiger |

Table 1: Top 10 actors returned by NR$^2$ algorithm

In order to prove that negative reinforcement helps in increasing diversity, we conducted experiments for various values of $\alpha$. Fig 6 shows the effect of negative reinforcement ($\alpha$) on density. As $\alpha$ increases, the density decreases i.e diversity increases. Similarly, Fig 4, 5 show that the country coverage and movie coverage increase with increase in $\alpha$. This is because, the negative mass from ranked nodes is propagated to a larger portion of the graph, thus picking the next node which is very far from the ranked nodes. A similar behavior is observed in case of $\beta$ too, as shown in Fig 7, 8, 9. With increase in importance to an absorbing node, positive energy is absorbed from unranked nodes, thus indirectly helping the negative energy propagation.

## 4.2 Text Summarization

In this section, we demonstrate the effectiveness of our algorithm on networks arising from text documents.

The input is a set of related documents and the task is to extract the important sentences to form a summary with no redundancy. In [4, 11], this problem is solved as a graph centrality problem. First, a similarity graph $G$ is constructed using sentences in the document. Stopwords are removed, and all the words are reduced to their root form through stemming. Sentences are considered as nodes and edges are set up based on cosine similarity. Finally, sentences are ranked based on PageRank technique and the top ranking sentences are picked to form the summary. The idea of the algorithm is that if a sentence is central, then it should be connected to many other importance sentences in the graph, and hence it must be important in the original document as well.

Though the algorithm succeeds in finding the important sentences, it fails to avoid redundancy in the top ranking sentences. This immediately leads to the inclusion of diversity in the ranking process so that a diverse set of sentences are extracted from the similarity graph. We show that our algorithm can be successfully applied to text summarization problem.

In our experiments, we used DUC 2002 dataset[3] to tune the parameters and DUC 2004 dataset to test the performance. DUC 2002 and DUC 2004 datasets consist of 60 and 50 document clusters respectively, with human written summaries for each cluster. The task is to generate a 100 word summary for each cluster reflecting the gist of input documents. To evaluate summaries, we used ROUGE (Recall Oriented Understudy for Gisting Evaluation) [8]. ROUGE generates recall value based on overlapping n-grams between human and system generated summary to evaluate its quality.

We followed the same graph construction procedure presented in [13]. We feed the sentence-position information as a prior distribution to Personalized PageRank (PPR), Grasshopper (GH), DivRank (DR) and our algorithm. If a sentence '$s$' appears at '$l$'th position in a document, then $r(s) \propto l^{-\gamma}$, where $\gamma$ is a positive parameter. Following [10], we used PageRank score (PR) as the 'relevance score' in MMR. For each algorithm, parameters were chosen to maximize the performance on the training set. In our method, we tuned $\lambda$, $\alpha$, $\beta$ and $\gamma$ in $[0, 1]$.

Table 2 shows the average ROUGE-1 score of different algorithms on the training set. Table 3 shows the average ROUGE-1 score of different algorithms on the test set. It is evident that our method works better than other greedy selection methods on the test set. Further, it is quite competitive to DivRank.

| Algorithm | ROUGE-1 | 95% C.I. |
|---|---|---|
| PR | 0.33005 | [0.31471, 0.34553] |
| MMR | 0.34763 | [0.33289, 0.36253] |
| PPR | 0.3458 | [0.33105, 0.36042] |
| GH | 0.34978 | [0.33652, 0.36298] |
| NR$^2$ | 0.34656 | [0.33299, 0.36002] |
| DR | 0.35054 | [0.33683, 0.36397] |

Table 2: Text summarization results on DUC 2002 dataset

| Algorithm | ROUGE-1 | 95% C.I. |
|---|---|---|
| PR | 0.36216 | [0.34944, 0.37572] |
| MMR | 0.37280 | [0.36062, 0.38489] |
| PPR | 0.38183 | [0.36865, 0.39366] |
| GH | 0.38722 | [0.37567, 0.39954] |
| NR$^2$ | 0.38810 | [0.37563, 0.40004] |
| DR | 0.39397 | [0.38193, 0.40681] |

Table 3: Text summarization results on DUC 2004 dataset

# 5. RUNNING TIME ANALYSIS

The running time of our algorithm is linear in $k$, because of the iterative behavior of the method. Our algorithm works particularly well when $k$ is small compared to the number of nodes in the graph (ex: k<100). This is because, we return only top-k nodes and do not compute the ordering for the rest of the nodes. On the other hand, DivRank computes the ranking for all the nodes in the graph in a single pass irrespective of $k$, which is a very costly process if the graph contains millions of nodes. Further, DivRank uses an approximation algorithm whose convergence is not guaranteed.

[3]Document Understanding Conference
http://duc.nist.gov/

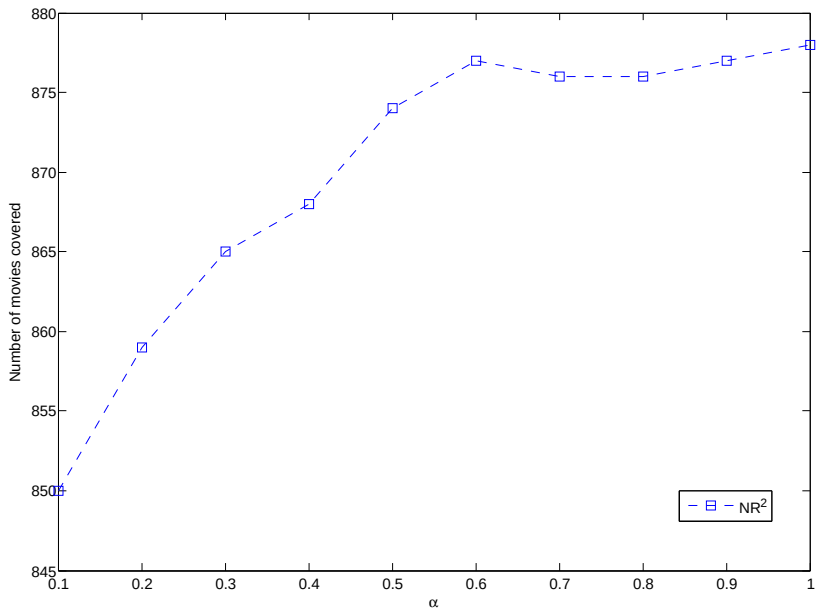

Figure 4: Movie coverage v.s. $\alpha$.

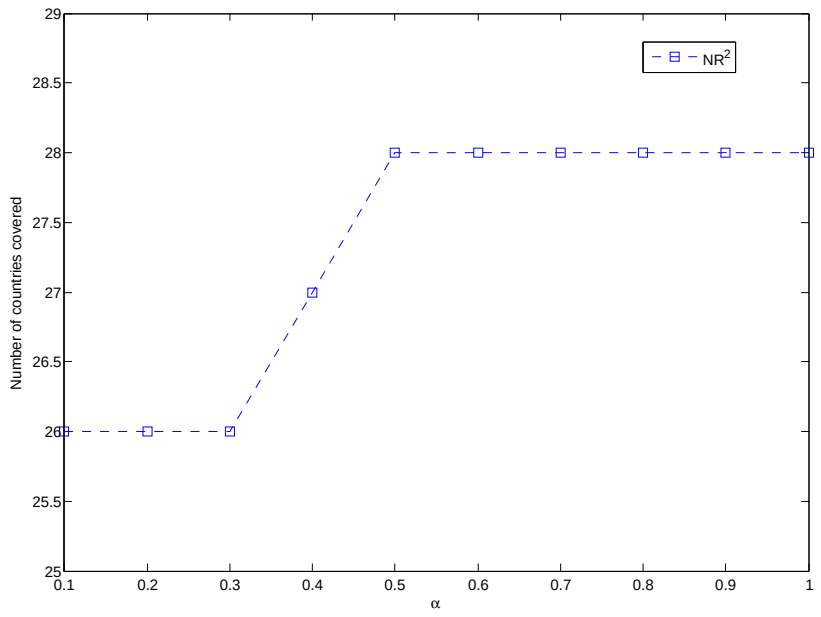

Figure 5: Country coverage v.s. $\alpha$.

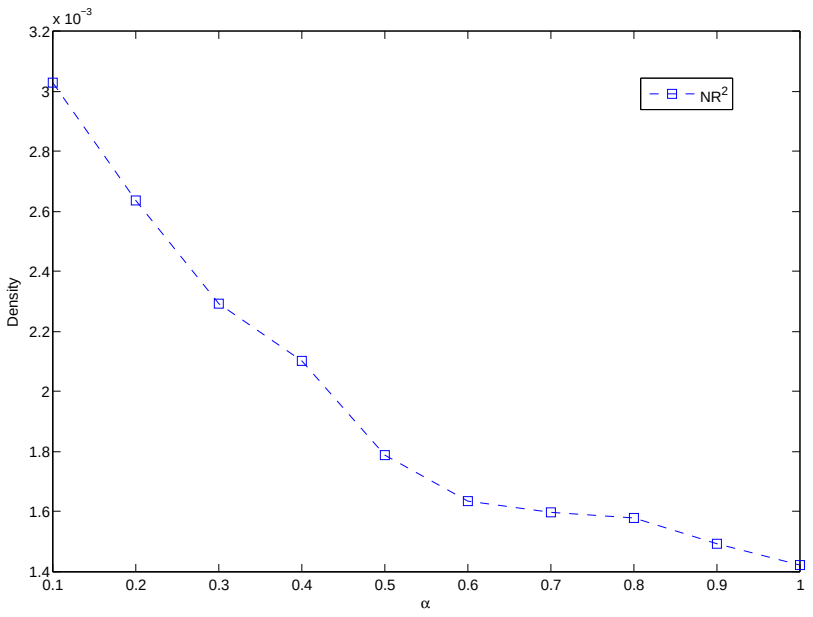

Figure 6: Density v.s. $\alpha$

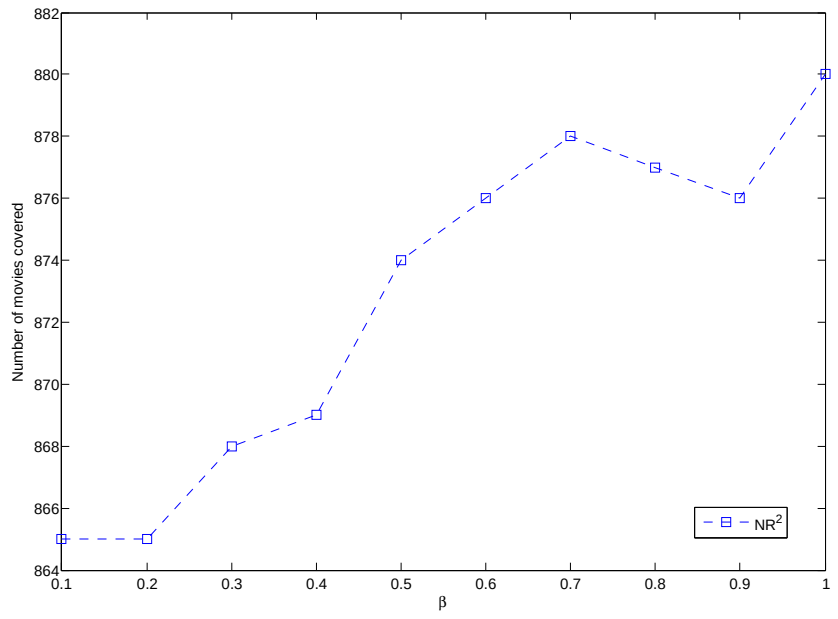

Figure 7: Movie coverage v.s. $\beta$.

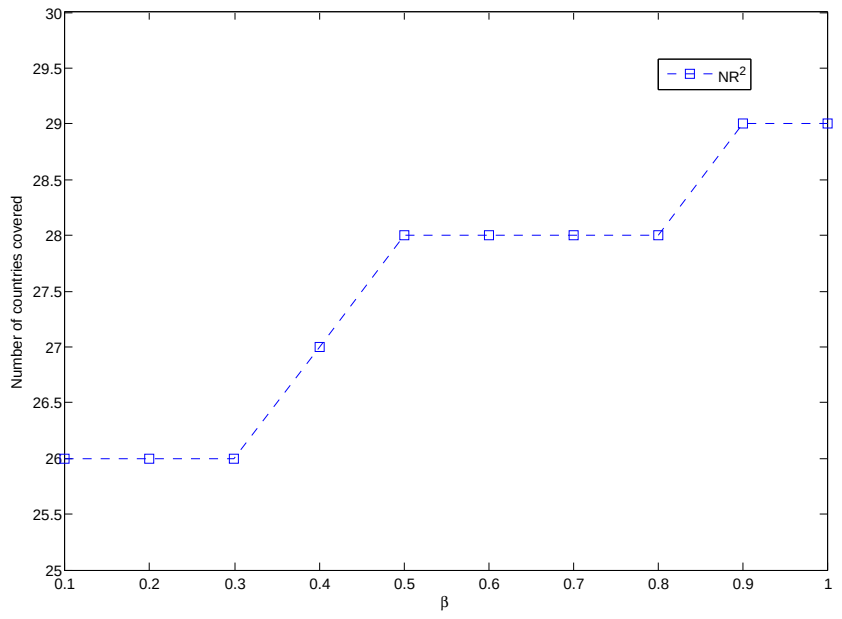

Figure 8: Country coverage v.s. $\beta$.

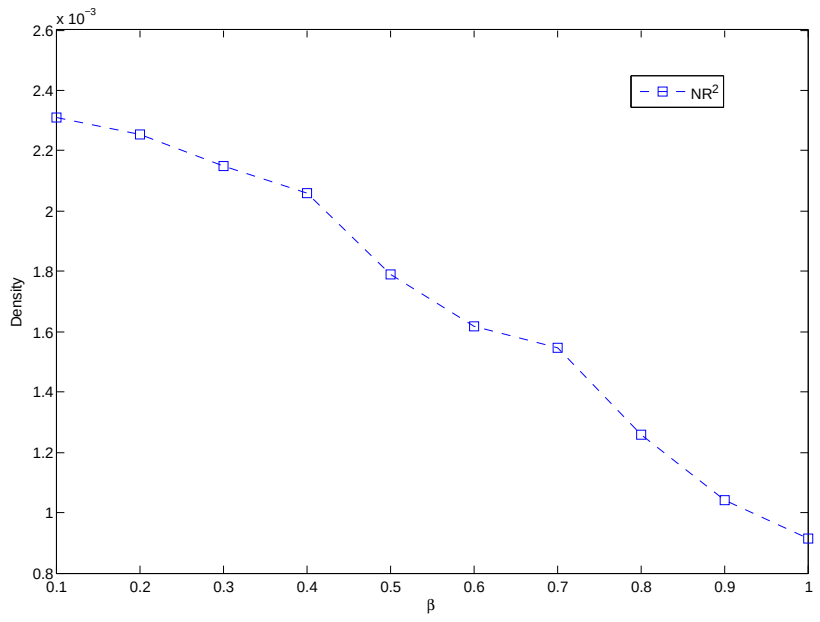

Figure 9: Density v.s. $\beta$

# 6. CONCLUSION

We have proposed a new diversity based ranking algorithm. Negative reinforcement is the core idea behind picking the next diverse item. Experiments on two benchmark datasets conclude that $NR^2$ is competitive to the existing methods. As part of future work, we are looking at how negative reinforcement can be utilized to turn our algorithm into a unified ranking one. The idea is to use both positive and negative reinforcement so that only the cluster centres benefit from positive reinforcement.

# 7. ACKNOWLEDGEMENT

We acknowledge a partial support for the work, from a project approved by the Department of Science and Technology, Government of India. We thank the reviewers for helping us improve the presentation.